\documentclass[superscriptaddress,twocolumn,showpacs,preprintnumbers,prl]{revtex4}
\usepackage{graphicx}
\usepackage{times}
\usepackage{epsfig}
\begin{document}

\def\salto{\vskip 1cm}
\def\lag{\langle}
\def\rag{\rangle}

\newcommand{\LANL} {Theoretical Division T-11, Los Alamos National Laboratory, Los Alamos, NM 87545, USA}
\newcommand{\ORNL} {Materials Science and Technology Division, Oak Ridge National Laboratory, Oak Ridge, TN 37831, USA}
\newcommand{\UT} {Department of Physics and Astronomy, The University of Tennessee, Knoxville, TN 37996, USA}
\newcommand{\UCSB} {Microsoft Project Q, The University of California, Santa Barbara, CA 93106, USA}
\newcommand{\UM} {Condensed Matter Theory Center, Department of Physics, The University of Maryland, College Park, MD 
20742, USA}

\title{Excitons in the One-Dimensional Hubbard Model: a Real-Time Study}

\author{K. A. Al-Hassanieh }   \affiliation {\LANL}
\author{F. A. Reboredo}           \affiliation {\ORNL}
\author{A. E. Feiguin }              \affiliation {\UM} \affiliation{\UCSB}
\author{I.  Gonz\'alez }              \affiliation {\ORNL} \affiliation {\UT}
\author{E.   Dagotto }                \affiliation {\ORNL} \affiliation {\UT}

\begin{abstract}
We study the real-time dynamics of a pair 
hole/doubly-occupied-site, namely a holon and a doublon, in a 1D Hubbard insulator 
with on-site and nearest-neighbor Coulomb repulsion.  Our analysis shows that 
the pair is long-lived and the expected decay mechanism to underlying spin excitations 
is actually inefficient.  For a nonzero inter-site Coulomb repulsion, we observe that part of the 
wave-function remains in a bound state.  Our study also provides insight 
on the holon-doublon propagation in real space.  
Due to the one-dimensional nature 
of the problem, these particles move in opposite directions even in the 
absence of an applied electric field.  The potential relevance of our results 
to solar cell applications is discussed.
\end{abstract}


\maketitle

{\it Introduction}.  In most band insulators and semiconductors, a single particle picture
is sufficient for a qualitative grasp of transport and optical
properties.  This simplicity was crucial for an early understanding of
their electronic structure and for discoveries, such as the
transistor and semiconductor solar cells. 
An equivalent progress for strongly correlated electronic
materials (SCEM), such as Mott-Insulators, has marched at a slower
pace since complicated many-body interactions govern their behavior,
and the one-electron picture breaks down [\onlinecite{Tokura}]. 
Nevertheless, gigantic optical nonlinear properties, potentially useful for applications, have been reported in 1D Mott-Hubbard materials, such as Sr$_2$CuO$_3$ [\onlinecite{Kishida}].
However, the majority of SCEM research has focused on fundamental science issues. 
While interesting technologies might emerge as a
result of the exotic transport properties and complex phase
diagrams of SCEM  [\onlinecite{Dagotto05}], their optical
properties have been largely regarded only as a probe of the ground
state. Whether SCEM can be of use for applications in solar
cells, photo-catalysis or solid-state-lighting devices will depend
on our basic understanding of their optical properties and
specifically on the dynamics of the optical excitations.
\vspace{0.0cm}

The market of light-to-energy conversion is currently dominated by
semiconductor materials, silicon in particular. Composite tandem
solar cells have been fabricated to take maximum advantage
of the solar spectra [\onlinecite{Bertness94}].  However, semiconductor solar
cells are very sensitive to defects in the crystal lattice, which
implies that cost remains a major limiting factor [\onlinecite{BES}].  Current efforts to reduce the cost of
solar cells include the exploration of other alternatives such as
composites of polymers, quantum dots, and their combinations
[\onlinecite{Kim07}]. In these novel devices, excitons
dissociate into electrons and holes at the interface of two materials
with different band offsets.
\vspace{0.0cm}

It is frequently claimed that to date ``all photo voltaic technologies
use semiconductor materials'' [\onlinecite{thormton02}].  While ideal 
solar cell materials have to fulfill a number of properties [\onlinecite{goetzberger03}], note
that having a gap between 1.1 and 1.7 eV, of {\it any} origin, is the
basic requirement. The fact that photo-catalytic processes are not quenched
in composites involving manganites [\onlinecite{shchukin99}]
raises hope for many highly correlated
oxides to be technologically useful for light-to-energy
conversion.  To control the values of their potentially useful intrinsic
gaps, transition metal oxides can be grown in complex layered
super-lattices [\onlinecite{Bhattacharya07,lowndes96}].  
\vspace{0.0cm}

However, one can argue against these highly correlated oxides as
candidates for solar energy harvesting materials since the optical gap
arises only as a result of electronic correlations: they would be metallic
otherwise. As a result, the ground state is magnetic with low-energy
excitations, which in principle can provide a path for the exciton decay.
Excitons in SCEM have received much theoretical attention 
[\onlinecite{Jeckelmann00,Gallagher97}],
but most of the efforts have focused on the exciton formation 
and its properties.  They have also been observed experimentally in 1D Mott insulators [\onlinecite{Ono}].  
A real-time study of the excitation propagation in real-space, accounting for its decay 
in addition to the bound state formation, has not been presented before to our knowledge. 
\vspace{0.0cm}

In this Letter, we study the dynamics of holon-doublon excitations 
in the 1D extended Hubbard
model using the recently developed time-dependent density-matrix renormalization group (TDDMRG). We 
find that: (i) The mechanism for exciton decay into
magnetic excitations is very inefficient and the pair holon/doublon is long-lived.  
This suggests that 1D SCEM can in principle be used to
generate power or promote chemical reactions at the surface, adding to its previously
discussed potential role as optical switches [\onlinecite{Kishida}];
(ii) at least in quasi-1D systems, despite the absence of an electric field, the holon and doublon 
move in opposite directions even in the presence of a finite attraction;  (iii) 
in agreement with previous calculations, a fraction of the pair forms a bound state.

{\it Model and Technique}. We investigate an open  Hubbard chain of $L$ sites 
with on-site and nearest-neighbor (NN) Coulomb repulsion. The number of electrons is 
set to $L$, i.e. the system is at half-filling. The Hamiltonian is given by
\begin{eqnarray}
\hat H \!&=&\! -t_h \!\sum_{\sigma, i=1}^{L-1} (c_{i\sigma}^\dagger c_{i+1\sigma} + H.c.) +  
U\!\sum_{i=1}^L (\hat n_{i\uparrow} - {1\over 2}) (\hat n_{i\downarrow} - {1\over 2})\nonumber \\
 &+& V\sum_{i=1}^{L-1} (\hat n_i - 1) (\hat n_{i+1} - 1), 
\end{eqnarray}   
where $t_h$ is the hopping integral, and $U$ and $V$ are the on-site
and NN Coulomb repulsion, respectively.  The rest of the notation is
standard.  The ground state $|\Psi_0\rangle$ of $\hat H$ is calculated
using static DMRG [\onlinecite{White92,Hallberg06}].  This state has a 
charge gap at $U\neq0$, and spin antiferromagnetic quasi-long-range 
order.    

\begin{figure}[t]
\epsfxsize=6.5cm \centerline{\epsfbox{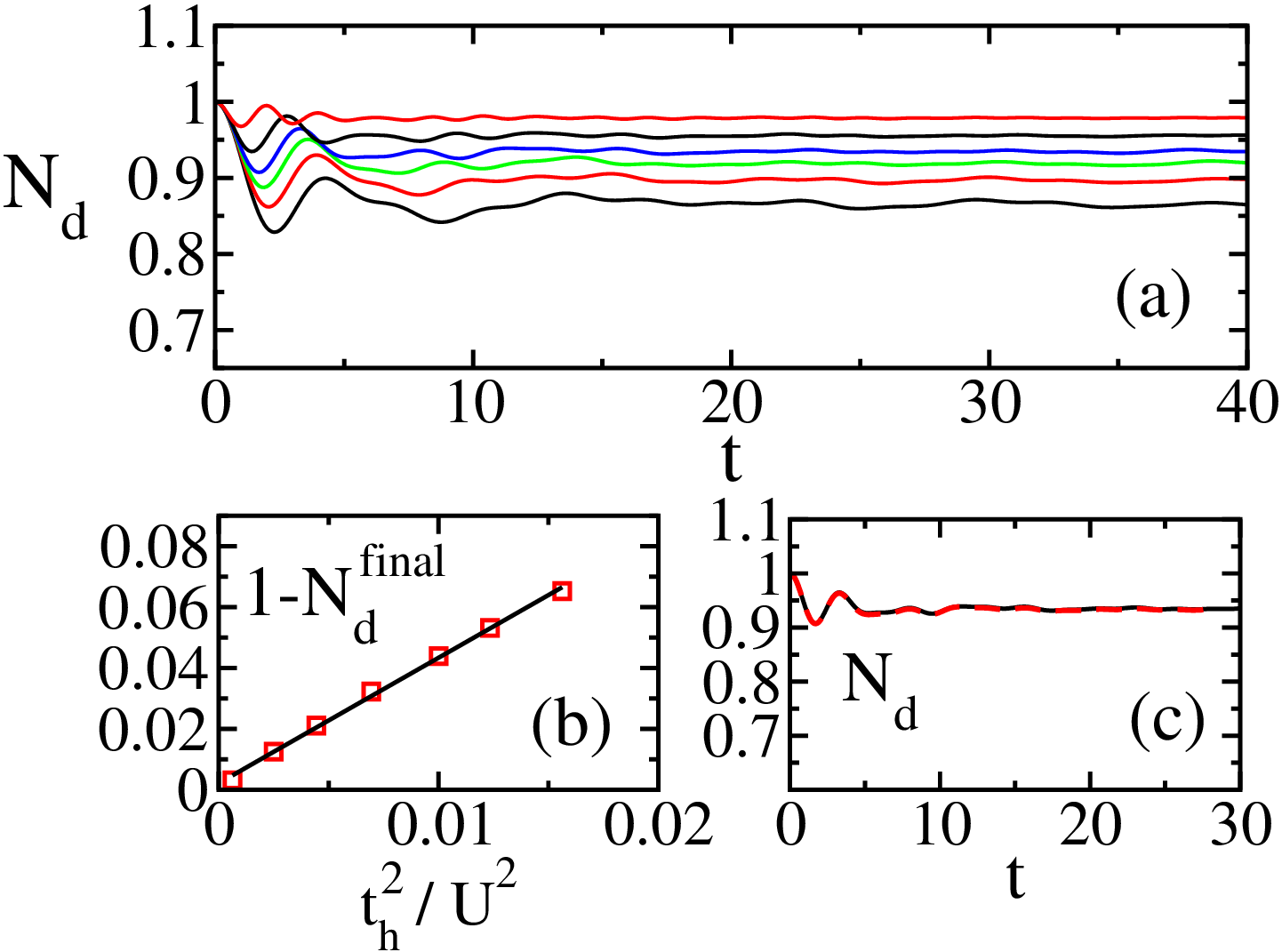}}
\caption{Decay of the holon-doublon excitation.  (a) Time-evolution of the total double occupation 
$N_d$ for $V=0$ and $U/t_h = 5, 6, 7, 8, 10$, and $15$ (bottom to top).  Note that even for moderate values 
of $U$, the decay is rather small (less than $15\%$ for $U=5t_h$).  (b) The recombined 
fraction, $1-N_d^{~\rm final}$, vs $U/t_h$ 
for $8t_h\le U \le 40t_h$.  $1-N_d^{~\rm final}$ scales as $t_h^2/U^2$.  For 
$U/t_h = 40$, the recombination is negligible and the system is in the 
strong coupling limit.  (c) $N_d$ for $U=8t_h$ and two chain lengths, $L=40$ (solid line) 
and $L=80$ (dashed line).  The results are independent of the system size.}
\label{Fig1}
\end{figure}

Light excitation of solids is a complex process in which an energetic
electron-hole pair is created by absorbing a photon. %%[\onlinecite{note4}].  
In general, the first pair created involves wave-functions not
included in the Hubbard model which only takes into account a narrow
window around the Fermi level.  In most solids, these ``hot'' electrons
and holes quickly dissipate energy into phonons until they reach the
quasiparticle energy minimum.  Having opposite charge, they find each 
other because of the long-range part of the Coulomb interaction.  We are 
not modelling the evolution of the state that
results from the initial absorption of light. Instead, our goal is to
investigate whether the electron and hole will recombine
non-radiatively when they find each other. We
model this situation by creating an excited state $|\Psi_e\rangle$
formed by a hole and a doubly-occupied-site on two neighboring sites
at the center of the chain: $|\Psi_e\rangle = {\it
  h}^{\dagger}_{L/2}{\it d}^{\dagger}_{L/2+1}|\Psi_0\rangle$, where
${\it h}^\dagger_i$=$(1/\sqrt{2})\sum_\sigma
c_{i\sigma}(1-n_{i\bar\sigma})$ and ${\it
  d}^\dagger_i$=$(1/\sqrt{2})\sum_\sigma
c^\dagger_{i\sigma}n_{i\bar\sigma}$ create a holon and a doublon,
respectively [\onlinecite{note0}].  Nearest-neighbors holon and doublon 
is the most favorable case for an eventual recombination.  This state does not correspond to 
the extended state that light creates but to one 
that can form after multiple collisions with the lattice.
$|\Psi_e\rangle$ is time-evolved under $\hat H$,
$|\Psi(t)\rangle$=$e^{-i\hat H t}|\Psi_e\rangle$.  Due to the third
term in $\hat H$, the holon and doublon experience an attraction $V$.
However, the system will try to equilibrate by delocalizing the
excitation via the kinetic energy term in
$\hat H$.  In the results shown, the measurements are done on
$|\Psi(t)\rangle$.  To time-evolve the many-body wave-function, the
Suzuki-Trotter (ST) decomposition version of TDDMRG is employed
[\onlinecite{White04,Daley04}].  Unless stated otherwise,
we use $L$=$40$, $M$=$200$ DMRG states, $t_h$=$0.2$, and the
ST time step $\tau$=$0.05$ [\onlinecite{note1}].

{\it Results}.  To study the decay of the excitation in real time, we follow the time-evolution of the
total double occupation $N_d = \sum_i n^d_i = \sum_i \langle \hat n^d_i \rangle = \sum_i \langle \hat n_{i\uparrow}
\hat n_{i\downarrow}\rangle$  [\onlinecite{note3}].  In the results for $N_d$, and the correlation 
function shown below, the corresponding ground state value is subtracted.  
Figure \ref{Fig1}(a) shows the time evolution
of $N_d$ for different values of $U/t_h$ at $V/t_h=0$.  A small
portion of the holon-doublon pair recombines at short times, while the remaining
fraction survives indefinitely [\onlinecite{note10}].  As $U/t_h$ increases, the holon-doublon recombination
becomes negligible, in spite of the availability of the spin excitations channel; 
in the strong coupling limit, the holon-doublon
pair is conserved.  Figure \ref{Fig1}(b) shows the asymptotic value of the 
recombined fraction of the pair.  For large $U/t_h$, this fraction scales
as $t_h^2/U^2$ or $J/U$, where $J$=$4t_h^2/U$ is the effective spin coupling.  This 
confirms that the recombination does involve spin excitations in the magnetic 
background.  However, while several spin excitations could be created 
leading to the decay of most of the holon-doublon pair, it is surprising 
that their 
number, proportional to the slope of the curve, is small.  This is mainly 
due to the 1D nature of the system, that allows for holons and doublons to freely
propagate even in a spin staggered background, 
as opposed to the well-known trapping of holes in 2D antiferromagnetic backgrounds due to
``string'' excitations [\onlinecite{Dagotto94}].  Figure \ref{Fig1}(c) 
shows $N_d$ for two different chain lengths with $L$=$40$
and 80.  The results are nearly identical indicating that the holon-doublon
recombination is independent of the system size.  Overall, 
Fig. \ref{Fig1} indicates that the mechanism of the
holon-doublon decay into spin excitations is inefficient, and a major
fraction of such a pair survives even for moderate values
of $U/t_h$.

\begin{figure}[t]
\epsfxsize=5.5cm \centerline{\epsfbox{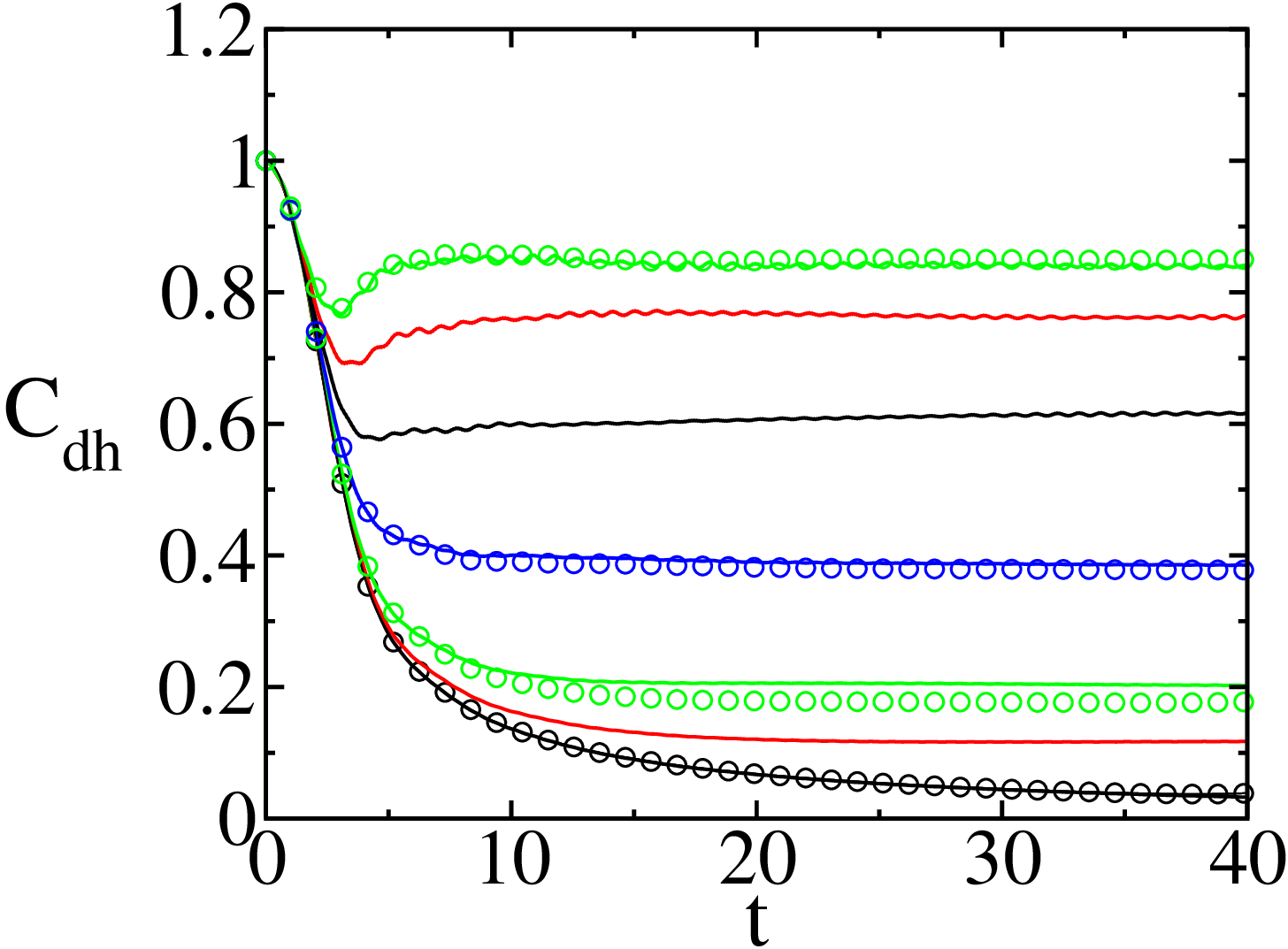}}
\caption{Time-evolution of the NN holon-doublon correlation 
function $C_{dh}$ in the strong coupling limit ($U=40t_h$) for $V/t_h=0, 0.5, 1.0, 2.0, 3.0, 4.0,$ and $5.0$ (bottom to top).  $C_{dh}$ 
saturates at a finite value, i.e. a bound state is formed, for any $V>0$.  The open circles show the 
NN charge-charge correlation function in the simplified model 
of two spinless fermions (see text for details).} 
\label{Fig2}
\end{figure}

It is also interesting to study the nature of the holon-doublon interaction,
in particular the formation of a bound state.  For this purpose, we
measure the NN holon-doublon 
correlation function $C_{dh} = \sum_i \langle 
\hat n^h_i \hat n^d_{i+1} + \hat n^d_i \hat n^h_{i+1}\rangle$.  Although this 
correlation function beyond NN can be finite (i.e. the 
average holon-doublon distance can be larger than $1.0$), 
a finite $C_{dh}$ indicates the formation of a bound state.  In the strong-coupling limit, the Hamiltonian 
reduces to an effective two-body problem, namely two spinless fermions 
on a tight-binding chain with NN {\it attraction} $V$.  % [\onlinecite{note2}].  
Previous studies 
in this limit [\onlinecite{Gallagher97}] suggest that a bound state 
at $K = \pm\pi$ is formed for any $V > 0$; whereas for $K = 0$, a bound state is formed 
only if $V \ge 2t_h$, where $K = k_d + k_h$ is the total 
wave-vector of the holon-doublon pair.  The exciton 
has the dispersion $E(K) = U -V - (4t_h^2/ V)\cos^2({K/2})$.  Our results 
agree with the above analysis.  However, since the two wave-packets contain  all possible 
wave-vectors, the  holon-doublon wave-function is expected to split into bound and 
continuum states depending on $V/t_h$.  
Figure \ref{Fig2} shows the time-evolution of $C_{dh}$ in the 
strong coupling limit for different values of $V/t_h$.  As expected for $V/t_h = 0$, 
$C_{dh}$ decays to zero at long times i.e. no bound state is formed. 
 For nonzero $V/t_h$'s, 
$C_{dh}$ shows a clear saturation.   
For large $V/t_h$, the 
saturation value approaches $1.0$, i.e. the holon and doublon remain bound at NN 
sites.  For comparison, an effective model of two spinless fermions on 
a tight-binding chain with NN attraction $V$ was also studied.  
The initial state, with the two particles at NN sites, 
is time-evolved exactly under the simplified Hamiltonian, and the 
NN charge-charge correlation function is calculated. 
The exact results of the simplified model (shown 
in Fig. \ref{Fig2}) 
are in very good agreement 
with the full model TDDMRG results.

\begin{figure}[t]
\vspace{0.0cm}
\epsfxsize=6.0cm \centerline{\epsfbox{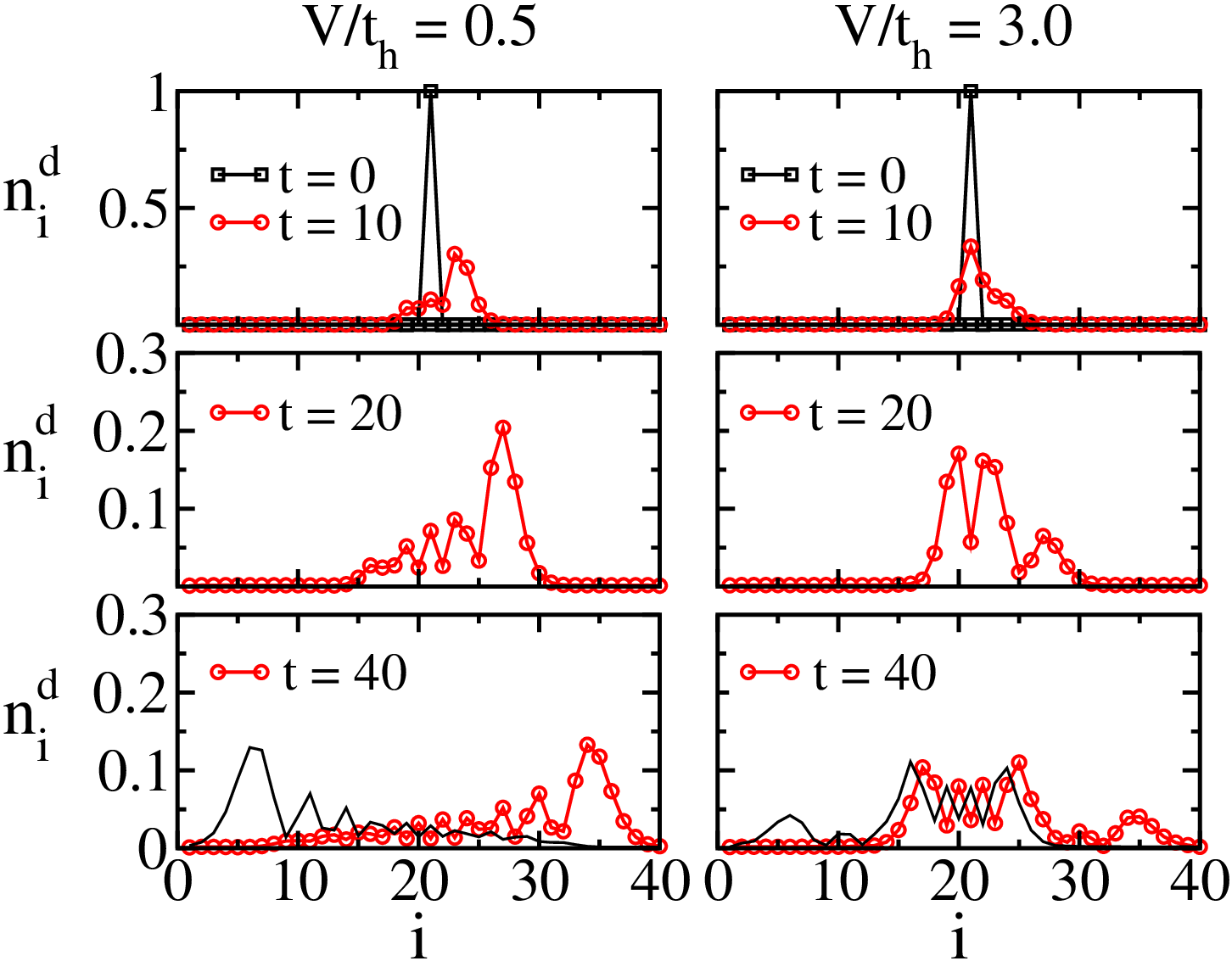}}
\caption{Doublon propagation in space, in the strong coupling limit ($U=40t_h$).  For $V=0.5t_h$, 
most of the doublon propagates to the right (i.e. most of its wave-function  
is in continuum states).  For $V=3t_h$, the major portion forms an excitonic  wave-packet.  
The transition from the small to the large $V$ behavior occurs at $V=2t_h$.  The holon motion, 
shown in the lowest two panels, is the reflection of the doublon motion, around the chain center.}
\label{Fig3}
\vspace{0.0cm}
\end{figure}   

Figure \ref{Fig3} shows the doublon propagation in real space.  For a small $V$, such as $V=0.5t_h$, 
most of the wave-packet propagates to the right [\onlinecite{note}], suggesting that a major fraction 
of the initial holon-doublon wave-function was in continuum states.  The motion of the holon  
is simply the reflection of the doublon motion around the 
center of the chain.  Note that, in principle, the holon and doublon should move symmetrically in 
both directions.  However, the movement to the left is prevented by energy conservation which 
forbids these particles from crossing each other via a virtual recombined state.  In 
other words, the holon acts as a hard-wall potential causing the doublon to move in one direction (similar
arguments hold for the holon motion).  
Note that, in addition to the dominant fraction that moves in one direction, there is 
a long tail of small weight that moves in the other direction.  This tail contains a mixture 
of continuum and bound states of finite $K$, and this behavior is observed for 
$0\le V < 2t_h$.  For larger $V$, such as $V=3t_h$, a smaller fraction of the doublon propagates 
to the right.  The dominant part forms a symmetric wave-packet that remains centered at the 
same site.  Results for other
values of $V/t_h$ show that the transition from the small to the large $V$ regimes 
occurs at $V=2t_h$ [\onlinecite{Khaled}].  Note that for $V>2t_h$, a bound state can be formed 
for any $K$.  Thus, the symmetric wave-packet observed is a mixture of bound states, 
with all values of $K$.  The bound and continuum states are clearly separated in space for $V>2t_h$.   
Even though $C_{dh}$ increases monotonically as $V/t_h$ is 
increased, note that the doublon propagation in real space falls into two distinct regimes of small 
and large $V/t_h$.  From the practical point of view, the amount of charge that can be collected 
at the ends of the chain is not affected by the bound state formation as long as $V<2t_h$.

 \begin{figure}[t]
\epsfxsize=6.0cm \centerline{\epsfbox{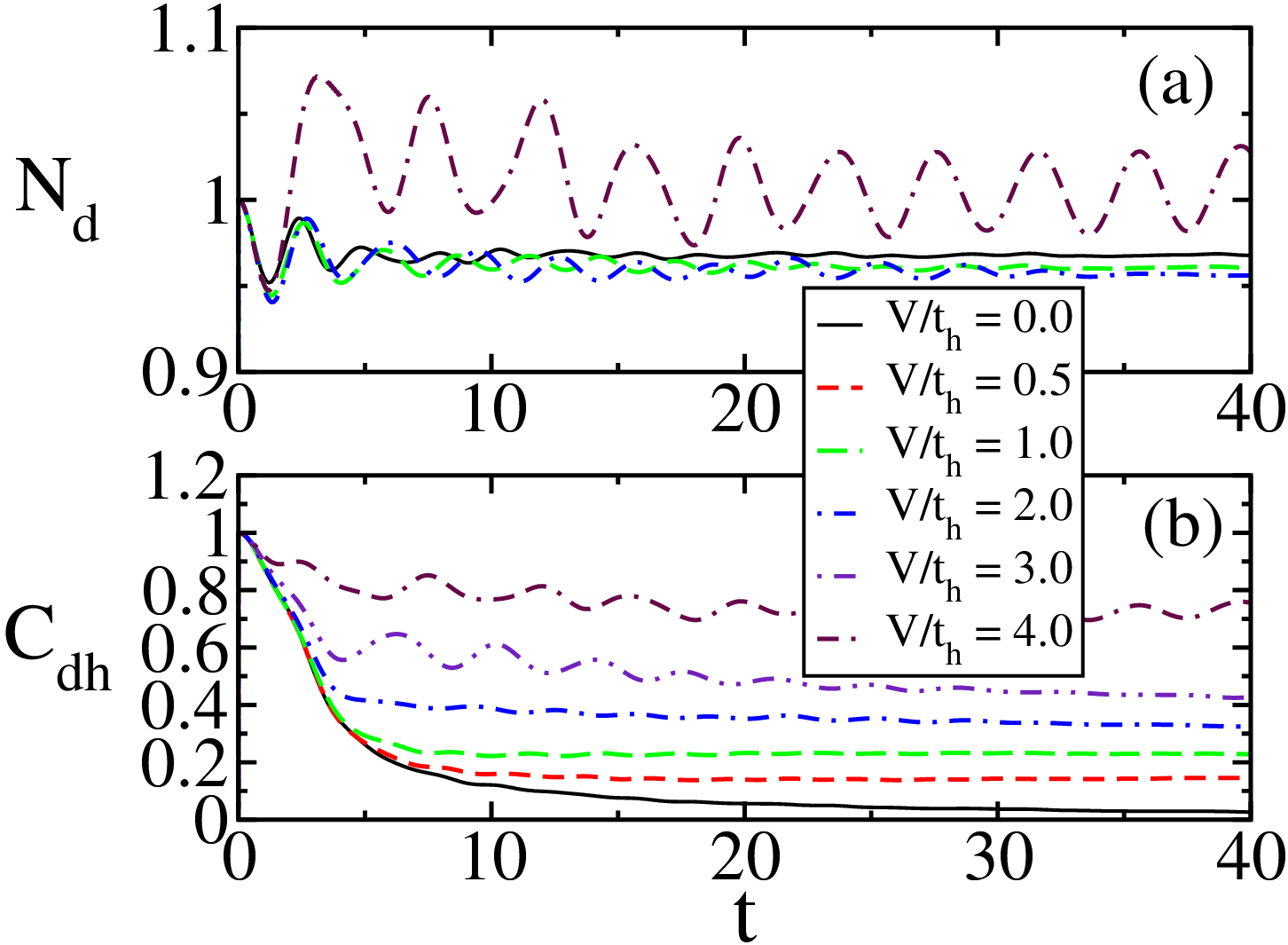}}
\caption{(a) $N_d$ and (b) $C_{dh}$ for $U/t_h$=$12$ and different couplings $V/t_h$.  
The final value of $N_d$ shows a nonmonotonic dependence with $V/t_h$ suggesting 
a competition between two tendencies.  Note, however, that the changes are small for 
the values of $V/t_h$ used.  $C_{dh}$ shows the same behavior as in the strong coupling 
limit.  Two plots were deleted from the upper panel for clarity. }
\label{Fig4}
\end{figure}

The results shown above
remain valid as couplings are reduced.  Figure 
\ref{Fig4} shows $N_d$ and $C_{dh}$ for $U/t_h$=$12$ and 
several $V/t_h$'s.  The values of $V$ are chosen such that the system is 
in the Hubbard insulator regime and far from the charge-density-wave (CDW) instability:  the 
transition between the two regimes occurs at $V$=$U/2$, 
i.e. $V$=$6t_h$ in this case [\onlinecite{Jeckelmann02}].  
The final value of $N_d$ shows a non-monotonic behavior as a function of $V/t_h$.  
This can be understood as the result of the competition between two tendencies.  On one hand, 
for a finite $V$ the energy added to the system upon creating the holon-doublon pair 
is $U-V$, to zeroth order in $t_h$.  That is, at large $V$ less energy is 
supplied to the system, thus $N_d$ can relax to smaller values.  On the other hand,  $V$ 
increases the CDW tendency in the ground state.  And as energy is supplied to the 
system (by creating the holon-doublon pair), the weight of the higher energy 
states (the CDW states in this case) increases, thus increasing $N_d$.  Note, 
however, that the effect of $V$ on $N_d$ is small 
as long as the system is in the Hubbard insulator regime.  $C_{dh}$, on the other hand, 
exhibits the same behavior as in the strong coupling limit: a bound state 
is formed for any $V>0$.   The holon and doublon propagation in space is qualitatively 
the same as in the strong coupling case (Fig. \ref{Fig3}).   

{\it Conclusion}.  We studied the time-evolution of a holon-doublon pair in a 1D extended Hubbard
model.  The naively expected decay mechanism 
of the holon-doublon pair to spin excitations is shown to be
inefficient, particularly at large $U/t_h$, and the pair survives 
for long times [\onlinecite{many-excitons}].  Depending on the value of the NN Coulomb repulsion, the 
holon-doublon wave-function splits into bound and continuum states.  Although a bound state can 
be formed for any finite NN repulsion, there is a qualitative 
difference in the real-space propagation between the weak and strong $V$ regimes.  
It is very important to remark that much of the above results arise from the 
1D nature of the problem.  The spin background affects much 
less the holon and doublon propagation in 1D than in higher 
dimensions [\onlinecite{Kim96}], 
where antiferromagnetic links 
are broken in the direction perpendicular to their movement creating ``strings'' 
[\onlinecite{Dagotto94}].  Thus, 
the decay by spin excitations of the holon-doublon pair in 1D is expected to be less likely than 
in higher dimensions.  For  these reasons, the main conclusion of our present theoretical effort is that
1D Mott-Hubbard insulators could potentially join semiconductors
and polymers as gapped materials of potential relevance for solar cell devices.

K.A. and F.R. thank C. Batista and I. Ivanov for useful
discussions.  K.A., F.R., I.G., and E.D. were supported by NSF grant
DMR-0706020 and the Div. of Mat. Science and Eng., U.S. DOE under
contract with UT-Battelle, LLC.  LANL is supported by US DOE under
Contract No. W-7405-ENG-36.


\begin{thebibliography}{99}
\bibitem{Tokura} Y. Tokura and N. Nagaosa, Science {\bf 288}, 462 (2000).
\bibitem{Kishida} H. Kishida {\it et al.}, Nature {\bf 405}, 929 (2000);
H. Kishida {\it et al.}, Phys. Rev. Lett. {\bf 87}, 177401 (2001);
H. Matsueda {\it et al.}, Phys. Rev. B {\bf 71}, 153106 (2005).
\bibitem{Dagotto05} E. Dagotto, Science {\bf 309}, 257 (2005).
\bibitem {Bertness94} K. A. Bertness {\it et al.}, Appl. Phys. Lett. {\bf 65}, 989 (1994).
\bibitem {BES} DOE BES Report ``Basic research needs for solar energy utilization,'' April 18-21, 2005. 
\url{http://www.sc.doe.gov/bes/reports/files/SEU_rpt.pdf.}
\bibitem{Kim07} J. Y. Kim {\it et al.}, Science {\bf 317}, 222 (2007); I. Gur {\it et al.}, Science {\bf 310}, 462 (2005); 
M. Law {\it et al.}, Nature Materials {\bf 4}, 455 (2005). 
\bibitem{thormton02} ``Photovoltaic System Design'' in {\it Encyclopedia of Physical Science and Technology}  Ed. R. A. Meyers, Academic Press (2002).
\bibitem{goetzberger03} A. Goetzberger {\it et al.}, Mat. Science and Eng. R  {\bf 40}, 1 (2002).
\bibitem{shchukin99}  
D. G. Shchukin {\it et al.}, Int. J. of Photoenergy {\bf 1}, 65 (1999).
\bibitem{Bhattacharya07} A. Bhattacharya {\it et al.}, Appl. Phys. Lett. {\bf 90}, 222503 (2007).
\bibitem{lowndes96} D. H. Lowndes {\it et al.}, Science {\bf 273}, 898 (1996). 
\bibitem{Jeckelmann00} E. Jeckelmann {\it et al.}, Phys. Rev. Lett. {\bf 85}, 3910 (2000);
K. Tsutsui {\it et al.}, Phys. Rev. B {\bf 61}, 7180 (2000);
F. H. Essler {\it et al.}, ibid. {\bf 64}, 125119 (2001);
E. Jeckelmann, ibid., {\bf 67}, 075106 (2003).
\bibitem{Gallagher97} F. B. Gallagher and S. Mazumdar, Phys. Rev. B {\bf 56}, 15025 (1997);
W. Barford, ibid. {\bf 65}, 205118 (2002); F. Gebhard {\it et al.}, Phil. Mag. B {\bf 75}, 47 (1997).
\bibitem{Ono} M. Ono {\it et al.}, Phys. Rev. Lett. {\bf 95}, 087401 (2005).
\bibitem{White92} S. R. White, Phys. Rev. Lett. {\bf 69}, 2863 (1992); Phys. Rev. B {\bf 48}, 10345 (1993).
\bibitem{Hallberg06} K. Hallberg, Adv. Phys. {\bf 55}, 477 (2006); U. Schollw\"ock, Rev. Mod. Phys. {\bf 77}, 259 (2005).
\bibitem{White04} S. White and A. Feiguin, Phys. Rev. Lett. {\bf 93}, 076401 (2004).
\bibitem{Daley04} A. J. Daley {\it et al.}, J. Stat. Mech.: Theory Exp. (2004), P04005.
\bibitem{note0} The definition of $h^\dagger$ and $d^\dagger$ assumes that the probability of double 
occupancy is small in the ground state.  This regime, intermediate and large $U/t_h$, is the focus 
of our effort here.
\bibitem{note1} $\hbar$ is set to 1 and time is shown in units of $1/5t_h$.
\bibitem{note3} The total hole occupation $N_h = \sum_i \langle (1-\hat n_{i\uparrow})(1-\hat n_{i\downarrow})\rangle$ 
is equal to $N_d$ due to the particle-hole symmetry of the model.
\bibitem{note10} The time scale of the recombination is roughly proportional to $1/U$.
\bibitem{Dagotto94} E. Dagotto, Rev. Mod. Phys. {\bf 66}, 763 (1994).
\bibitem{note}  The velocity of the doublon is what we expect for the used $t_h$.
\bibitem{Khaled} K. A. Al-Hassanieh {\it et al.}, in preparation.
\bibitem{Jeckelmann02} E. Jeckelmann, Phys. Rev. Lett. {\bf 89}, 236401 (2002).
\bibitem{many-excitons} For many excitons, the recombination could occur by the transfer of energy 
from one exciton to another.  The efficiency of this mechanism will be tested in future investigations.
\bibitem{Kim96} C. Kim {\it et al.}, Phys. Rev. Lett. {\bf 77}, 4054 (1996).

\end{thebibliography}
\end{document}